\documentclass[10pt,journal,compsoc]{IEEEtran}
\usepackage[utf8]{inputenc}
\usepackage{adjustbox}

\usepackage{flushend}

\begin{document}

\title{How Widely Can Prediction Models be Generalized?\\ An Analysis of Performance Prediction in Blended Courses}
\author{ Niki Gitinabard, Yiqiao Xu, Sarah Heckman, Tiffany Barnes, Collin F. Lynch
}

%
%

\markboth{IEEE Transactions on Learning Technologies}%
{Shell \MakeLowercase{\textit{et al.}}: Bare Demo of IEEEtran.cls for Computer Society Journals}

\IEEEtitleabstractindextext{%
\begin{abstract}
Blended courses that mix in-person instruction with online platforms are increasingly popular in secondary education. These tools record a rich amount of data on students’ study habits and social interactions. Prior research has shown that these metrics are correlated with students’ performance in face to face classes. However, predictive models for blended courses are still limited and have not yet succeeded at early prediction or cross-class predictions even for repeated offerings of the same course.

In this work, we use data from two offerings of two different undergraduate courses to train and evaluate predictive models on student performance based upon persistent student characteristics including study habits and social interactions. We analyze the performance of these models on the same offering, on different offerings of the same course, and across courses to see how well they generalize. We also evaluate the models on different segments of the courses to determine how early reliable predictions can be made. This work tells us in part how much data is required to make robust predictions and how cross-class data may be used, or not, to boost model performance. The results of this study will help us better understand how similar the study habits, social activities, and the teamwork styles are across semesters for students in each performance category.  These trained models also provide an avenue to improve our existing support platforms to better support struggling students early in the semester with the goal of providing timely intervention.
\end{abstract}

}

\maketitle
\IEEEraisesectionheading{\section{Introduction}\label{sec:introduction}}
The use of technology and online tools in undergraduate courses is expanding rapidly and blended courses are becoming the norm in postsecondary education. The tools that are used in these courses include: learning management systems (LMSs) such as Moodle and Canvas which are used to distribute course materials; discussion forums such as Piazza which are used by students to seek help or to collaborate with others; automated submission and grading systems such as WebAssign for assignments and automated feedback; and more recently software development tools such as Github or Jenkins which support realistic assignments and help prepare students for their future jobs. In addition to supporting students' learning, these tools provide us with rich data on the students' online behavior and study habits as well as their social connections, performance, and help-seeking. 

Until recently rich detailed data of this kind has been rare in education and has only been available in Massive Open Online Courses (MOOCs) and other purely online courses. Student behaviors in MOOCs has been studied extensively. This research has shown that students' social interactions and online behaviors on MOOCs can be used to predict their performance as well as the likelihood that they will complete or drop out of the course \cite{joksimovic15,hmelo14, yang13, kovanovic14, eckles12, jiang14, ramesh13, deboer13, yang15, yang13, crossley2015language, tucker2014mining}. Prior research on MOOCs has also shown that the prediction models were applicable on data from other offerings of the same course, different courses, and even using only data from early weeks of those classes  \cite{brooks15, jiang14b, taylor14, dekker2009predicting, halawa2014dropout, brinton2015mooc}. 

Researchers have begun to mine the rich online data from blended courses to develop predictive models that can be used to understand students' habits and to predict their performance (e.g. \cite{watson14, vihavainen13, zafra2011, ibrahim2007predicting}).  In prior work, for example, we evaluated features of students' study habits and social interactions on some of our current blended datasets \cite{gitinabard17w, sheshadri18}.  We found that these features could be used to predict the students' performance on the same offering.  While these studies have been informative, most of the work has been focused on analyzing a single offering of a course (e.g. \cite{zafra2011}) or on replicating findings from one course in another (e.g. \cite{andres18}), not on developing models that can be used across classes or class offerings.

This is crucial because in order for a model to be useful it must be the case that we can train it before it is actually necessary.  If we have to wait until a given class offering is wholly complete and we have the outcomes before we can train a model then any guidance based upon it would be not useful. Therefore, in order for such predictive models to be useful, we must be able to train them on one class or class offering and use them on another.  Thus, they must be able to rely on features of the students, such as their study habits and social behaviors, that can persist across classes, and it must be the case that we can make reliable predictions early in the course, while there is still time for the students to choose a better path.


Our goal in this work is to address this issue by developing cross-class models of student performance based upon students' study habits and social relationships. Unlike performance metrics which may be specific to a course or to an assignment, we hypothesize that positive and negative study habits will be persistent for students across classes and as a result, that predictive models based on these habits should generalize from one class to the other. In this study, we will use data collected from two offerings of two different introductory CS courses at NC State University to address the following research questions:

\begin{itemize}
    \item \textbf{RQ1.} How do different methods of social graph generation affect the performance of predictive models based upon them? 
    \item \textbf{RQ2.} What features of students' study habits and social connections are most predictive of student performance?
    \item \textbf{RQ3.} How early can we predict students' performance in these classes using the data from the same class?
    \item \textbf{RQ4.}  Will predictive models generated from one offering of a course transfer to another offering of the same course?
    \item \textbf{RQ5.} Will prediction models generated on one course transfer to another?
    \item \textbf{RQ6.} How will these models perform in identifying at-risk students?
\end{itemize}
This work will highlight the potential for predictive models based upon real-time data extraction from different online learning platforms to provide guidance to students during a course.

\section{Background}

\subsection{Social Learning \& Study Habits}

We chose to focus on models based upon social relationships and study habits in light of prior research showing that these features are both essential to learning, and generally persist across classes.  Social learning has been studied by a number of researchers including Bandura \cite{bandura77} as well as Lave and Wenger \cite{lave91}.  Bandura, in particular, noted that learning through social connections and by example is often more efficient than learning through direct experience and practice.  He argued that successful social learning was driven in part by the students' ability to connect with and to reproduce the lessons of their peer group.  Lave and Wenger articulated a complete theory of learning as an inherently social activity.  Professionals, small communities, and experienced groups constitute \emph{communities of practice} (e.g. professional programmers, computer scientists, or students in CSCS226) with their own distinct group knowledge and practices  \cite{lave91,Wenger98}.  In their formulation, students are novices who seek to learn this knowledge and move into these communities by emulating members of the group or even co-constructing new knowledge and engaging in ``legitimate peripheral participation'', by engaging in practices that emulate professional activities, and by absorbing shared practices.  Their ability to do so properly depends upon their social connections and the extent to which they seek help from others. Lave and Wenger further argue that these communities not only help students learn course material better but also engaging in these communities can give the students some sense of belonging, thus the students who participate more are less likely to leave the course.  Other researchers have studied this framework in online platforms before (e.g. \cite{johnson01, borthick2000, fisher98, hammond98}). In most of this prior work, the researchers concluded that in order for the students to work towards solving a problem as a community, it is better for them to have face-to-face interactions. Thus, they propose some online platforms for these discussions to take place and they argue that when these platforms are good enough, the students' experience will be sufficiently close to what face-to-face interaction feels like and thus closer to traditional communities of practice. However, there is some debate about how well the theory of communities of practice fits into online discussion forums in MOOCs.  In blended courses, the discussion forum is only an addition to the students' interaction modes.  They can still meet and know each other in person and engage in other richer forms of interaction.  They are also more likely to have other preexisting social relationships with one-another which they will bring to class.  Thus, we hypothesize that the social interactions recorded should be closer to full social learning and thus more effective. Additionally, the students' use of the discussion forums is determined by their individual study habits and social behaviors that are more general than a single class.  That is why we believe that these features will be robust in the face of differences in class structure and content and are appropriate for use in cross-class evaluation.

Prior research has also investigated different individual programming and study patterns among students and observed that the better performing students usually have distinguishing habits. Often, these studies are based upon snapshots of the students' code and recordings of their activities, investigating how these habits correlate with the students' performance. Several researchers have noted that lower performing students usually take longer to complete the exercises \cite{ahmadzadeh05, uchida02, spacco15}. Prior research that has tracked student activities during their coding sessions has also shown that better performing students usually analyze the code in a more logical manner and spot the issues faster, while lower performing students have difficulty localizing problems \cite{uchida02, lin16}. Aside from observing students while coding, other researchers have inferred students' coding behaviors from online activities or code snapshots recorded by the IDE \cite{carter15, blikstein11,chao16, watson14, vihavainen13, hosseini17, brinton2015mooc}. They have shown that students with some general online habits such as those who prefer to spend more time on the lecture videos, or who pausing more than once, or rewind at least once are more likely to perform better in class \cite{brinton2015mooc}. As these studies show, better-performing students tend to have online behavioral patterns that distinguish them from the lower-performing groups and most of these habits do not seem to be class-specific. Thus, we choose to also look into the students' online habits and patterns of work to make predictions both within and across classes.

\subsection{Predictive Models in MOOCs}

All course activities in a typical MOOC occur online.  Students view lecture videos, complete assignments electronically, and seek help via a single platform or a suite of tools that are all controlled by the course provider.  And as the students are spread across the globe and are largely unknown to one-another, this platform is typically the only option for communication.  This creates data chokepoints that have yielded a wealth of data on students' performance, communications, and study habits (e.g. data from Coursera and EdX platforms \cite{gitinabard18, andres18, xumany}). This data has been used to support a great deal of work on predictive and analytical models both within- and across-classes.

\subsubsection{Within Class Predictions}

While MOOCs are quite popular, they are also characterized by high levels of dropout sometimes as much as 90+\% \cite{Rivard:Measuring:2013}.  As a consequence, prior researchers have focused heavily on predicting not only how students will perform but who will actually finish, or dropout. Some have used metrics extracted from the students' social interactions and their social presence \cite{joksimovic15, hmelo14, yang13, kovanovic14, jiang14, jiang14a, brown15, rose14, zhu16}. These metrics include measures of connectivity, engagement, and online presence which can be calculated from their interactions.  This work has been driven by the premise that students who are more engaged with their peers are both more likely to learn and less likely to quit. Joksimovic et al., for example, analyzed students' social presence through features such as continuing a thread, complimenting other users, and expressions of appreciation \cite{joksimovic15}.  They found that these metrics can be used to predict the students' final grades on the course. Yang et al. applied survival analysis to identify the most important features for predicting the students' dropout \cite{yang13}. The analyzed features included those of students' posting behavior such as post length, students' enrollment date (students frequently join MOOCs after the official start date), and social network features.  Ultimately they concluded that the date students enroll in the course, their post length (i.e. how long in average their posts are), and their authority score (i.e. how much the reply to their peers who ask a lot of questions) were the most informative of the available features. Kovanovic et al. \cite{kovanovic14} and Jiang et al. \cite{jiang14} also analyzed students' social network features.  Both groups found that centrality, the extent to which a student is on the shortest path between other individuals, was predictive of their performance.  Kovanovic et al. also found that  the students' interactive social presence (e.g. whether or not they were affective, interactive, or cohesive) was highly correlated with their other social metrics and their performance.

Other researchers have analyzed informative metrics based upon students' general study habits such as: the number of posts, time spent online, when they joined the course, the number of videos watched in a week, the number of quizzes or assignments they attempted, the number of forum posts made per week along with the post length, the time spent on assignments, whether they spend more time on forums or on the assignments, whether or not they start early, and other demographic data such as their age, fluency with English, and their education level \cite{ramesh13, deboer13, yang15, yang13, pursel16, fei15, rose14, andres18, chen17, sinha14}. These features are typically defined with the goal of capturing common behaviors, such as starting assignments early or writing more detailed posts, among better performing groups and using those to classify students by performance or persistence. While most of these features can be calculated directly from the students' user-system interaction logs, some researchers have gone further to analyze generated features such as study sessions, action sequences and estimates of confusion, which must be constructed from groups of logs and which require some estimation and analysis \cite{yang16, chen17, brooks15, kloft14, amnueypornsakul14, li17, sinha14}. These features are intended to capture complex behavioral patterns among students that can highlight reasons for failing or dropping out of a course such as general confusion, dissatisfaction, or boredom. 

Many of the more complex models included study sessions which were generated from a sequence of student interactions with the system during a set span of time. Amnueypornsakul et al. for example, set study sessions and used the features extracted from those sessions such as the length of the sequence, the number of occurrences, and the number of Wiki page views to train a predictive model for attrition using Support Vector Machines. However, their results don't seem to be promising as their F1 score was approximately 0.2 \cite{amnueypornsakul14}. Li et al. defined sessions as well, but applied N-gram classification techniques from natural language processing to the action sequences to predict whether or not the students would obtain a certification of the students.  They assumed that the better performing students usually take specific sequences of actions that can distinguish them from lower performers \cite{li17}. They used Logistic Regression models and were able to achieve an F1 score between 0.5 and 0.6. Brooks et al. defined different fixed-duration sessions which ranged from one day to an entire semester, and then classified the students' level of activity in each window as a binary feature (active vs. inactive) \cite{brooks15}. They further defined sequences over these values and then classified them using n-grams to predict dropout.  These sequences can be used to show whether or not a student has long periods of inactivity or whether they work diligently.  They combined these features with a decision tree model and were able to predict Distinction group of the students ($grade>85$) with a $\kappa$ of 0.9 or above for different classes. Their high performing prediction models can show that most of the high performing students are identifiable by their amount of activities during different course timeframes. Sinha et al., by contrast, defined a network over the students and course resources, connecting each student to the online resources they accessed.  They then used metrics collected from that network to predict student performance and were able to outperform the model based on N-grams of students' activities \cite{sinha14}. They used cost sensitive LibSVM with radial basis kernel function (RBF) as the learning algorithm and were able to achieve an accuracy of about 0.6 and a $\kappa$ around 0.3 for different configurations. The graph metrics on their kinds of network shows the frequency of the students' access to different class material and tools and the findings of their study show that in their case, this frequency has been more informative than the sequences of actions. 

Other researchers also focused on the content generated by the students and used text-based features to make predictions of the students' performance \cite{crossley2015language}. Crossley et al. for example extracted linguistic features and applied a multivariate analysis of variance(MANOVA) for statistical analysis. They found that the average post length, the word age of acquisition (i.e. the age at which a word is typically learned) for words in the post, the use of Cardinal numbers, Hypernymy standard deviation (Hyponymy shows the relationship between a generic term and a specific instance of it, this measure shows how specific or generic their language is), Situational cohesion,  and Trigram frequency are helpful measures when predicting student performance as they can show the relative complexity of the students' posts which in turn can relate to the amount of time they spent writing the content \cite{crossley2015language}. Then they used these features for a Stepwise Discriminant Function Analysis and were able to predict student retention within the course with an accuracy of 67.8 and a $\kappa$ of 0.379.

\subsubsection{Cross Class Predictions}
While many features have been used successfully for within-class performance prediction, they do not always generalize across classes, due in part to variations in course structure or content.  Nor do they always generalize even to different offerings of the same class, due in part to variations in the student populations.  Some of the features described above, however, do generalize and can be used to make predictions based upon a few weeks of data \cite{boyer15, brooks15, jiang14b, taylor14, dekker2009predicting, halawa2014dropout, brinton2015mooc, Wei2017convolution, he2015identifying}. Boyer et al., for example, identified two different kinds of learning models: models trained on the entire history of the class, and models trained on a moving time window of the class \cite{boyer15}. They also specifically considered transfer learning, bringing information from a prior course to make predictions in an ongoing one. They found that the performance of an a-posteriori model, based on the entire history of a class, is more accurate than a predictive model based on a real-time moving window.  They concluded that this was due to the fact that the real-time model did not have access to all of the necessary information for accurate prediction, but that its accuracy was more realistic. They used data from  three offerings of a class and trained a logistic regression model on the two earlier offerings before testing it on the first four weeks of the last one. They achieved an Area Under the Curve (AUC) score of 0.6-0.7 by the end of week four using models from previous offerings. Brooks et al. showed that by using data from the first three weeks of the class they could reach a moderate accuracy ($\kappa>0.4$) when identifying high performing students (students with a grade higher than 85) \cite{brooks15}. They also showed that using models trained on the first two offerings of a class to make predictions on a third offering can be done with moderate accuracy ($\kappa > 0.5$). Jiang et al. were also able to make predictions of  students' performance using a regression model and features such as social network degree, the number of completed assessments, and the average quiz score in the first week. By using data from the end of the first week they were able to identify students who achieved distinction in the course with an accuracy of 92.6\% \cite{jiang14b}.

\subsection{Predictive Models on Blended Courses}
Unlike MOOCs which offer rich and relatively comprehensive datasets, blended courses are far more challenging to analyze particularly given how much of the students' interactions (e.g. classroom lectures and direct peer contact) are not captured.  As a consequence, research on these courses has been more limited and far fewer analytical methods have been tried.  Moreover, most of this research has been focused on making within-class predictions using data from the entire semester. Watson et al., for example, defined features based upon the students' programming behavior in an introductory programming course to teach Java to students of varying abilities \cite{watson14}. Some examples of these features are the time students spent resolving a specific type of error or the frequency with which they transitioned between different types of errors and between error states and success. They then used a regression analysis to predict the students' performance.  They found that predictions based upon the student's observed programming behaviors were the most informative.  This is not entirely surprising given the close association between coding and introductory coursework.  Ibrahim et al. used general information on the students such as their knowledge of information technology applications, previous school type (boarding or non-boarding),  general programming knowledge, and family financial status to predict the students' undergraduate Cumulative Grade Point Average (CGPA) \cite{ibrahim2007predicting}. They used methods such as Decision Trees, Linear regression, and Artificial neural networks in their models. The average Root Mean Squared Error (RMSE) for all of their models was around 0.2 while Artificial neural network was the best performing of all. Zafra et al. analyzed students' activities on a learning management system (LMS) and utilized features such as number of assignments completed in the course, total time spent on the assignment section, number of messages posted to the forum, number of messages read on the forum, total time spent on the forum, number of quizzes seen, number of quizzes passed, number of quizzes failed, and total time spent on the quizzes to predict the students' final performance \cite{zafra2011}. They used several machine learning algorithms such as Sequential minimal optimization, Naive Bayes, Rep Tree (i.e. a type of decision tree implemented in Weka which minimizes the total variance of numeric features \cite{weka}), Decision Stumps (a decision tree with a depth of one \cite{weka}), and Multiple-instance logistic regression.  With these models they were able to achieve an accuracy of $\approx 0.7$ and a specificity of $\approx 0.6$. When analyzing a smaller online course, Macfadyen et al. defined features based upon the students' activities on the course LMS such as the total number of discussion messages posted, the total number of mail messages sent, and the total number of assessments completed. Using these features and a logistic regression model they were able to identify 81\% of the failing students by the end of the semester and overall 73.7\% of their predictions were correct \cite{Macfadyen2010MiningConcept}. In their study, they showed the potential for LMS data to make early predictions of the students' performance and using those predictions to issue early warnings. Additionally, Vihavainen et al. showed the potential of applying models across classes by training a non-parametric Bayesian network using B-Course on data from an undergraduate programming course and applied it to a synchronous math class \cite{vihavainen13}. Their study, however, was done on a single cohort of students who they tracked across classes and is not affected by the differences between the students of different classes. 


\section{Dataset Information}
\begin{figure*}[h]
  \centering
  \includegraphics[width=0.77\textwidth]{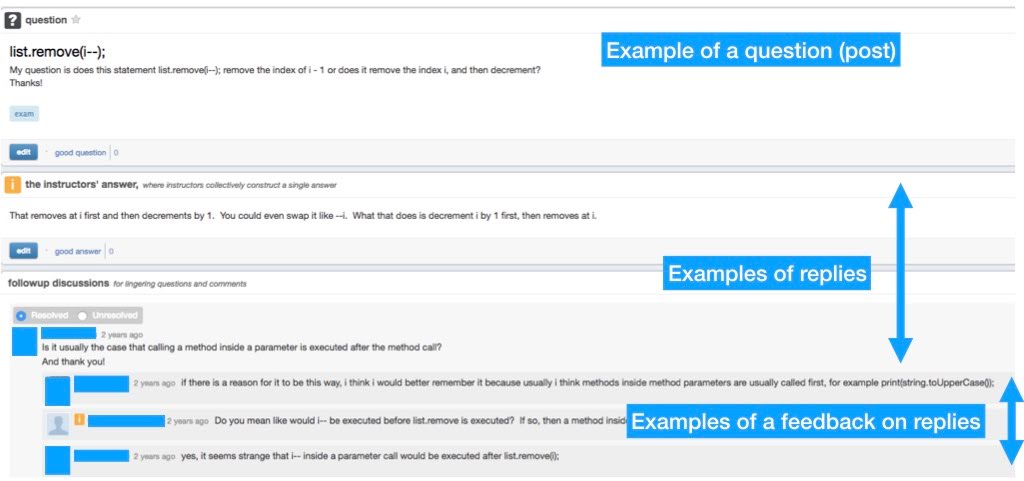}
  \caption{An example of a Piazza Question and the Followups}
  \label{fig:piazza_example}
\end{figure*}

Our analyses in this work is based on two offerings each of two distinct courses ``Discrete Mathematics for Computer Scientists'' (DM) which was collected in the Fall semesters of 2013 and 2015, and ``Java Programming Concepts'' (Java) which was collected in the Fall of 2015 and 2016. DM-2015 and Java-2015 occurred contemporaneously. Both DM and Java are core undergraduate courses that are required for all students majoring in Computer Science. Students typically take both courses during their second or third year in school and among the students in our analysis 126 are common between DM-2015 and Java-2015. Our analysis in this work does not include information on the students from the other classes they were taking. The main reason for not adding this information is generality of the analysis. We want this work to be replicable on other classes and the data from other simultaneous courses is not necessarily available, as in DM-2013 and Java-2016. Both courses use significant online materials and support including online assignments, supplemental material, and student forums. Thus they are prototypical blended courses. Both classes use the Piazza discussion forum. Piazza is structured as a question-answering platform. Students open a thread by asking questions or making a posts. The students and instructor can then reply to the question or reply to the replies. An example thread is shown in Figure \ref{fig:piazza_example}.  As the figure illustrates, the responses from the instructors and TAs are flagged as ``instructor answer'' and are distinct from the student responses. In each reply, the students can post a feedback or ask more questions and keep a discussion going. Instructors can also explicitly mark a post as `resolved' after they reply to it. Students can also recommend or upvote posts as needed. In some courses, posts may be made anonymously with the identity being kept secret from other students (partial anonymity) and possibly from the instructors as well (complete anonymity). The completely anonymous posting was permitted briefly in DM-2013 but was blocked in all other courses while partially-anonymous posting was always allowed. Participation in Piazza was highly encouraged, though not mandatory, in all of the courses. 

The topics covered in DM include: propositional logic, predicate calculus, methods of proof, elementary set theory, the analysis of algorithms, and the asymptotic growth of functions. The main focus of the Java class is on software system design and testing; encapsulation; polymorphism; composition; inheritance; linear data structures; specification and implementation of finite-state machines; interpretation of inductive definitions (functions and data types); and resource management. Some information on the population of these courses is shown in Table~\ref{tab:stats}. Our datasets consist of the Piazza discussions, Moodle logs, and final grades for all the classes as well as Github commit logs for Java classes and WebAssign logs for DM-2013.

\begin{figure}[h]
  \centering
  \includegraphics[width=0.47\textwidth]{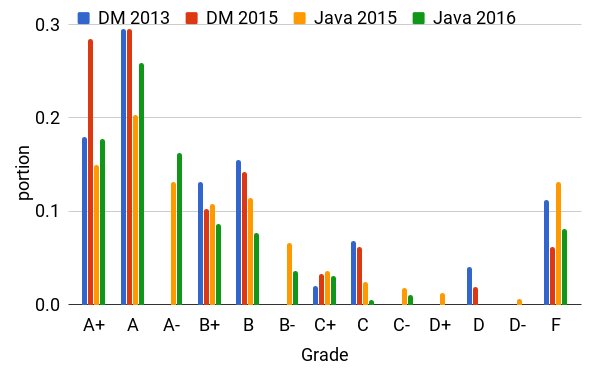}
  \caption{The Distribution of Grades in Different Classes}
  \label{fig:grade_dist}
\end{figure}

The grade distributions for these classes are shown in Figure \ref{fig:grade_dist}. As this figure shows, most of the students performed well in the classes. Thus, we concluded that partitioning them into pass/fail groups would be uninformative and result in a skewed dataset. Since the median grades for all these datasets were close to 90 which is the cutoff between an A- and a B+ in the course, we decided to segment the students into A- or above and B+ or below. We, therefore, partitioned the classes into two groups, the \emph{distinction} group who earned an A- or above, and the \emph{non-distinction} who earned a B+ or below.  This cut-off value resulted in an almost even partition of the students. We believe that this segmentation leaves room for adjusting the analysis for other classes with different grade distributions.

\begin{table}
\centering
\caption{Statistics of Each Class}
\label{tab:stats}
\begin{adjustbox}{width=0.48\textwidth}

\begin{tabular}{|l|c|c|c|c|} \hline
\textbf{Class} & \textbf{DM-2013} & \textbf{DM-2015} & \textbf{Java-2015} & \textbf{Java-2016}\\ \hline
Total Students & 251 & 255 & 181 & 206 \\ 
Teaching Assistants & 5 & 5 & 9 & 9 \\
Instructors & 2 & 2 & 4 & 4 \\
Average Grade & 81.2 & 87.6 & 79.7 & 79.9\\
\hline
\end{tabular}
\end{adjustbox}
\end{table}

\subsection{Discrete Math Classes}

A total of 251 students enrolled in DM-2013, while DM-2015 had a total of 255 students. In both semesters, the class was offered in two sections taught by the two instructors with 5 shared teaching assistants. The average final grade in DM-2013 was 81.2 and 87.6 in 2015 class. Both sections in each offering used the same Moodle webpage for sharing assignments, a Piazza forum for discussions, and both used WebAssign alongside hand-graded homeworks.  In these classes, the students achieved a 90\% or better average on the first three assignments (which were completed three weeks in the course) were offered a role as a peer tutor.  Peer tutors who completed ten hours of scheduled support for their classmates, by holding in-person office hours or answering questions online, were permitted to skip the final exam. The only substantive structural difference between the two courses was that in 2015 the instructor consciously delayed responding to posts on Piazza so that the TAs and peer tutors would be more involved.  Most of the posts were still answered in the same amount of time with the lead TA providing most of the responses.

\subsection{Java Programming Concepts Classes}

A total of 181 students completed the Java course in 2015 while the 2016 class had a total of 206 students. In both years, the course was offered in two different in-person sections with two separate instructors as well as a distance education section with two instructors, for a total of four different instructors with shared teaching assistants. We ignored the distance education students in our analysis because they were a much smaller group and differed substantially from the local students who can engage in face-to-face interactions.  However, they are included in our social network structures as they replied to questions by other students on the same forum.  These classes used Piazza for discussions, Moodle for sharing course materials, Github for working on group projects, and Jenkins for automated code evaluation.

\section{Methods}

For our analysis, we first extracted individual social networks from the courses and defined browser and study sessions to group students' online activities. We then extracted suitable quantitative metrics from these structures and used those metrics to train predictive models. As discussed above the models were trained on a single class offering and were then evaluated both within and across classes. 

\subsection{Defining Social Networks}

There are a number of different ways to extract social networks from online student interactions. The variation in these methods is due to differing assumptions that are made about the unrecorded student behaviors or about the `meaning' of the digital records. Several of these have been previously explored with MOOCs.  In this research, we focused on two distinct methods described by Brown et al. \cite{brown15,brown15w}, and Zhu et al. \cite{zhu16}. We designate these methods A and B respectively.  In both methods each node in the graph represents a forum participant while each directed arc represents a communicative relationship.   

In method A, based upon prior work by Brown et al. \cite{brown15, xumany} we make the assumption that everyone who replies to a thread has read all of the prior posts and replies within it before making their contribution.  Therefore, we connect the author of every reply to all the authors who contributed to the thread including the head post. Thus, in method A, a directed edge (u, v) is defined between users $u$ and $v$ for each instance where $u$ replied to a thread later than $v$. Our prior work showed that this method gives the best results when predicting student performance in MOOC forums where opening a thread shows you all the replies without needing to click on them \cite{gitinabard18, xumany}.

\begin{figure}[h]
  \centering
  \includegraphics[width=0.35\textwidth]{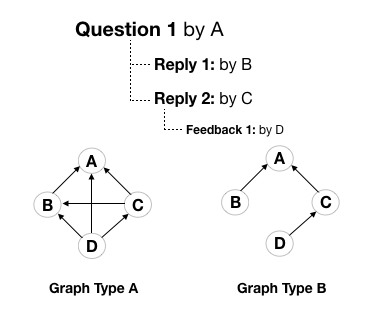}
  \caption{An Example of the Two Different Methods of Graph Generation}
  \label{fig:graph_types}
\end{figure}

In method B, based upon the work by Zhu et al. \cite{zhu16}, by contrast, we only assume that any author is replying to the head post of the thread and to any reply that they specifically respond to if any.  We do not make connections to the other replies. Under this more conservative assumption the network is limited to \emph{explicit} social connections rather than the additional \emph{implicit} ones incorporated above. Thus, in method B, a directed edge (u, v) is created between users $u$ and $v$ if $u$ has replied to a post by $v$ or commented on a reply by them. Our prior work on MOOCs has shown that when a forum structure requires the participants to click on every reply to read it completely and the replies are shown in short forms, method B works better than method A \cite{gitinabard18}. An example of these two methods of graph generation is shown in Figure \ref{fig:graph_types}. 

The structure of the student forums and their relative size is markedly different in blended courses and MOOCs. Most of the threads produced in our blended courses are shorter with an average length of 1-2 posts and replies, when compared to an average length of 5-6 in MOOCs with some MOOC threads reaching as long as 90 posts. In this part of our study, we generated both types of graphs to assess whether these assumptions affect the sensitivity and reliability of our predictions.  Each node in these graphs represents a participant in the course (\emph{Instructor}, \emph{TA}, or \emph{Student}). In the DM-2013 class, posting completely anonymously was allowed. This produced unknown author posts and replies which were removed from the analysis. The graphs include all  student interactions with other peers or with the teaching staff. We then aggregated the links between each pair of users to produce a single directed arc that was weighted by the number of communicative arcs in each direction. 

\subsection{Study and Browser Sessions}

In our prior work \cite{sheshadri18}, we sought to analyze students' study habits within courses by analyzing their study \emph{sessions}.  To that end, we collected log data from all of the different online platforms used in the DM-2013, DM-2015, and Java-2015 courses and unified them into a single transaction log.  We then used a data-driven method to segment this log into individual study sessions and analyzed the students' behaviors within these sessions.  We found that the properties of these sessions were significantly different among different performance groups and they can be used to make predictions on the students' performance.

We applied this same technique to generate unified logs for our current dataset.  These logs included 285,465 total actions from the  DM-2013 class, 24,180 actions from the DM-2015 class, 135,351 actions from the Java-2015 class and 175,059 from the Java-2016 class. Most of these transactions were WebAssign actions from DM-2013 and Moodle actions from the other courses.  The large divergence in total actions between the DM-2013 and DM-2015 datasets is due to the fact that the bulk of the transactions in the dataset are WebAssign activities which were not available to us in 2015.

In order to effectively analyze how students work we needed to group these individual actions into coherent sequences or sessions.   Grouping these actions is a nontrivial problem and, as Kovanovic et al. \cite{Kovanovic:Penetrating:2015} argued, how this grouping is defined can substantially affect the outcome of any analysis.  Fixed time durations have been used in prior studies (e.g. \cite{Brooks2015}) but this method can artificially separate actions that otherwise occur together (e.g. assignment submissions at 12:00 am and 12:01) and thus seemed inappropriate for our task where some students chose to work in short bursts and others regularly pull all-nighters.  Likewise, methods based upon collecting browser histories or additional data were intractable given our inability to access that information.  Therefore we selected a cutoff time between actions based upon our existing dataset.  This approach is similar to the one taken by  Amnueypornsakul et al. \cite{amnueypornsakul14} where we group all the actions that are within ``m'' minutes from each other in the same session and as soon as an action is later than m minutes from the previous one, we assume that a new session has begun. We defined those cutoffs based on our data and the general trends we could observe in the students' behavior \cite{sheshadri18}. We defined two different cutoff times (m) since the types of sessions can be different based on what resources are used and whether or not the students switched platforms. WebAssign and Moodle, for example, record clickstream actions which are comparatively brief, while Piazza records posting and editing questions and replies which takes a longer time. Additionally, longer breaks between actions can indicate the students going offline and working on a problem on paper or reading class material. 

Based on the properties of our data and the patterns of students' work, we decided to define the following session types with different cutoff values:
\begin{itemize}
    \item \textbf{Browser Session: } $m=15$ minutes indicating a short break is likely with the same browser open.
    \item \textbf{Study Session: } $m=40$ minutes indicating that the student likely changed tasks or quit working entirely. 
\end{itemize}

\emph{Browser Sessions} can be viewed as times when the students worked on a single task. This may include students working on multi-part WebAssign questions, reading through materials on Moodle, or working through an issue with their code with guidance from Piazza. Sessions of this type are comparatively short in duration. The \emph{Study Session} by contrast allows for larger gaps where students may shift from reading materials to answering questions or engaging in (online) discussions with their peers, and back again. This large cutoff was based in part upon the cross-platform breakdown and was in part intended to address our lack of data regarding the students' offline activities.  

\subsection{Feature Extraction}
We extracted two different classes of features from the social networks and browser sessions described above.  We calculated these features at three different time-points during the semester, before the first mid-term, before the second mid-term, and at the end of the class.  These differing cutoffs were used to determine how early we would be able to reflect on the students' performance. We discuss these features in greater detail below.  

\subsubsection{Features based on the Social Graph}

We calculated the following social metrics: 
\begin{itemize}
    \item \textbf{In-degree} shows the number of replies and feedback the student has received.
    \item \textbf{Out-degree} indicates the number of replies and feedback the student has given
    \item \textbf{Betweenness Centrality} is defined as a measure of the extent to which a vertex lies on the shortest path between others \cite{freeman}. Betweenness centrality tells how important this user is in connecting different users to each other, nodes with high betweenness are described as having some degree of control over the communication of others. \cite{freeman}
    \item \textbf{Hub and Authority Scores} are defined as mutually reinforcing scores: a good hub is a node that points to many good authorities; a good authority is a node that many good hubs point to \cite{kleinberg}. Users with high hub scores are those who frequently respond to the other active learners that post questions on the forum as the students with high authority scores are the ones that receive the most replies from the hub students.
\end{itemize}

\subsubsection{Features based on Sessions}
Our previous study of the sessions showed that many of the features that we defined were correlated with the students' performance \cite{sheshadri18}. We therefore chose the following metrics when making predictions about the student performance. We generated the metrics for the browser sessions and study sessions separately, to determine if one of these session types was more informative than the other.  The features are as follows:
\begin{itemize}
    \item \textbf{Number of sessions}: How many separate times the student has gone online throughout the semester, or in a specific timespan
    \item \textbf{Average number of actions in sessions}: How many actions the student usually gets done every time they start a session
    \item \textbf{Total number of actions}: How many total actions the student has done throughout the semester or during a specific timespan
    \item \textbf{Average duration of sessions}: How long the student usually stays online every time they start a new session
    \item \textbf{Overall Time spent in sessions}: How much time approximately the student has spent accessing online class tools, calculated as $time=Number\_of\_sessions \times Average\_duration\_of\_sessions $
    \item \textbf{Average gap between sessions}
    \item \textbf{Inconsistency}: How different the number of the sessions started by a student is from class average and how infrequent they get online, calculated as $Inconsistency= Average\_Gap \times (max(Number\_of\_sessions) -Number\_of\_sessions)$
    \item \textbf{Number and proportion of Homogeneous sessions} (where students were active on one platform only): In how many sessions and what proportion of sessions has the student focused on the same platform
    \item \textbf{Number and proportion of Heterogeneous sessions} (where students switched platforms): In how many sessions and what proportion of sessions has the student switched platforms
    \item \textbf{Ratio of sessions containing Piazza activity}: In what proportion of sessions has the student made a post or a reply on Piazza
    \item \textbf{Number of Piazza questions}: Number of posts made by the student
    \item \textbf{Number of Piazza answers}: Number of replies made by the student
\end{itemize}

\subsection{Performance Prediction}
As discussed before, we focused on classifying the students into two groups: \textit{distinction} (A- or above) and \textit{non-distinction} (B+ or below). We then divided the performance prediction step into three different rounds. In the first round, we trained and tested the machine learning models on the same class using 5-fold stratified cross validation and recorded the average F1 score over the 5 rounds of tests. These predictive models will show us how well those features are able to predict the students' performance and how early they can be used to make a prediction.

In the second round, we trained a predictive model on the earlier offering of each class (DM-2013 and Java-2015) and then tested them on the later offerings (DM-2015 and Java-2016) of the same class to see if these predictive models are generalizable from one offering to the other.

In the third round of our analysis, we applied the predictive model that we trained on the DM-2013 class (i.e. the earliest offering), on both offerings of the Java class which were offered after that, to see if the predictive models could be generalized across later offerings of different courses as well.

We used three classes of models: Random Forests, Support Vector Machines (SVM), and a Logistic Regression.  All were trained using the Scikit Learning library for Python \cite{scikit-learn}.  All of the models were tuned using a 5-fold cross validation via a grid search to fit the best parameters for the data. We chose these models because they are both fast and interpretable in this context. The random forest model was tuned to find the best max depth, the logistic regression was tuned on its penalty, tolerance, and \emph{C} (i.e. inverse of regularization strength where smaller values show stronger regularization.), and the SVM model was tuned to find the best values for \emph{C} and $\gamma$ for the RBF model (i.e. how far the influence of a single training example reaches, low values meaning far and high values meaning close \cite{scikit-learn}) in each fold. Finally, we evaluated the models according to their average F1 score using stratified 5-fold cross validation. F1 score or F measure shows the average between precision and recall \cite{scikit-learn}. In our case, it evaluates the model based on the proportion of the distinction students the model can find and what proportion of the guessed distinction students were true.

\section{Results and Discussion}

\subsection{RQ1. How do different methods of social graph generation affect the performance of predictive models based upon them?}
\begin{figure}[h]
  \centering
  \includegraphics[width=0.37\textwidth]{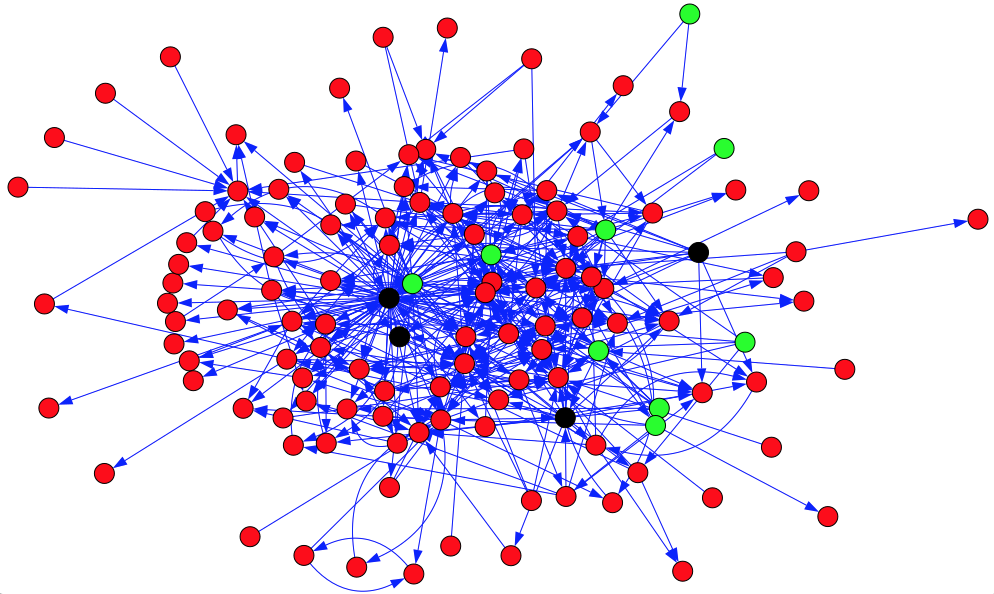}
  \caption{Graph generated using method B ($G_B$) for the Java-2015 class}
  \label{fig:java_2015_type2}
\end{figure}

To answer our first research question, we generated both types of graphs for all the classes and compared their effectiveness in making performance predictions. Two examples of the graphs are shown below. Figure \ref{fig:java_2015_type1} shows the graph that was generated using method A ($G_A$) from the Java-2015 dataset while Figure \ref{fig:java_2015_type2} shows the graph that was generated using method B ($G_B$) with the same class. Here the red nodes are the students, the green nodes are teaching assistants, and the black nodes are the instructors. As we would expect, while the main structure of these graphs are similar, $G_A$ has more edges than $G_B$,  as in generating $G_A$ we add one for each pair of posts in a thread, in each thread we connect the author of every reply to all the authors before, compared to $G_B$ where the reply authors only get connected to the main post author. As a result, $G_A$ is a proper subgraph of $G_B$, many of the links in $G_A$ are absent from $G_B$ and many others have less weight.

\begin{figure}[h]
  \centering
  \includegraphics[width=0.37\textwidth]{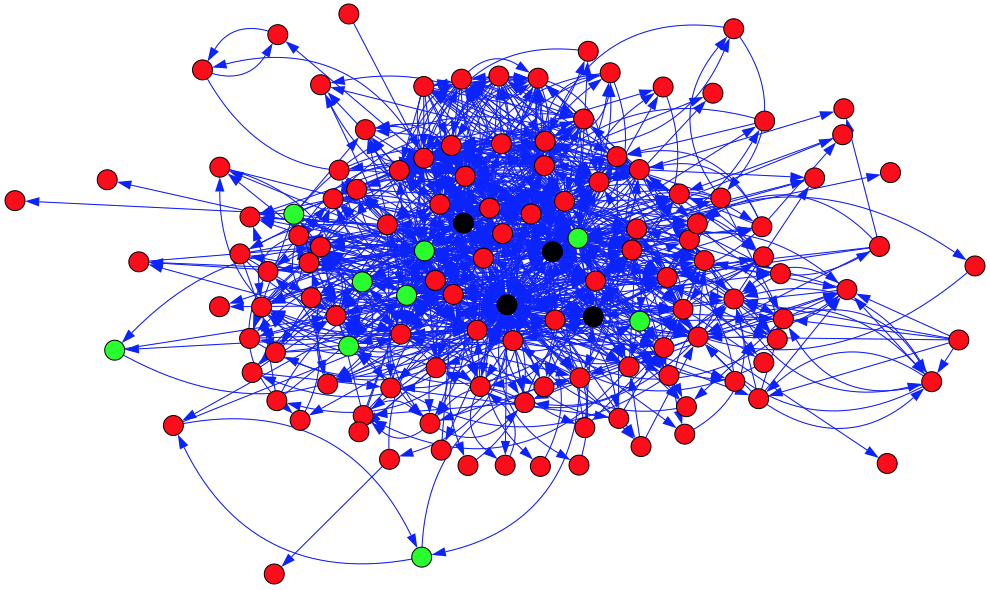}
  \caption{Graph generated using method A ($G_A$) for the Java-2015 class}
  \label{fig:java_2015_type1}
\end{figure}

In order to evaluate these graphs, we calculated the correlation between the graph attributes and the students' class performance.  Since our grades and graph attributes are not normally distributed, we chose to apply the nonparametric Spearman's Correlation Coefficient. The correlation coefficients and the p-values are reflected in Table \ref{tab:graph_corr}. Our results show that, consistent with our expectations, most of the graph attributes correlate with the students' performance more strongly in $G_A$ which connects the replying users to all the users who posted and replied to the same question earlier than them.  While in-degree, betweenness centrality, hub score, and authority score have higher correlations with the students' performance in $G_A$ for all the classes, the only exception is the out-degree which has a higher correlation in $G_B$. However, the correlations between out degree and performance are very similar in $G_A$ and in $G_B$ and the difference between them is usually very small. 


\begin{table*}[]
    \centering
    \begin{tabular}{|l|l|cc|cc|cc|cc|cc|}
    \hline
Class Name & & \multicolumn{2}{c|}{In} & \multicolumn{2}{c|}{Out} & \multicolumn{2}{c|}{Betweenness} & \multicolumn{2}{c|}{Hub} & \multicolumn{2}{c|}{Authority} \\
& & \multicolumn{2}{c|}{Degree}& \multicolumn{2}{c|}{Degree} & \multicolumn{2}{c|}{Centrality} & \multicolumn{2}{c|}{Score} & \multicolumn{2}{c|}{Score} \\
& & $\rho$ & P-value & $\rho$ & P-value & $\rho$ & P-value & $\rho$ & P-value & $\rho$ & P-value\\
\hline
 DM-2013 &  &  &  &  &  &  &  &  &  &  & \\
 & Before Test 1 type A& 0.320 & 0.000 & 0.347 & 0.000 & 0.313 & 0.000 & 0.339 & 0.000 & 0.313 & 0.000\\
 & Before Test 1 type B & 0.239 & 0.000 & 0.345 & 0.000 & 0.212 & 0.001 & 0.345 & 0.000 & 0.219 & 0.000\\
 & Before Test 2 type A& 0.338 & 0.000 & 0.367 & 0.000 & 0.339 & 0.000 & 0.367 & 0.000 & 0.349 & 0.000\\
 & Before Test 2 type B & 0.261 & 0.000 & 0.396 & 0.000 & 0.237 & 0.000 & 0.394 & 0.000 & 0.252 & 0.000\\
 & Full Course type A& 0.421 & 0.000 & 0.423 & 0.000 & 0.410 & 0.000 & 0.430 & 0.000 & 0.423 & 0.000\\
 & Full Course type B & 0.323 & 0.000 & 0.425 & 0.000 & 0.280 & 0.000 & 0.416 & 0.000 & 0.312 & 0.000\\
 \hline
DM-2015 &  &  &  &  &  &  &  &  &  &  & \\
 & Before Test 1 type A& 0.253 & 0.000 & 0.289 & 0.000 & 0.199 & 0.001 & 0.296 & 0.000 & 0.293 & 0.000\\
 & Before Test 1 type B & 0.234 & 0.000 & 0.292 & 0.000 & 0.171 & 0.003 & 0.298 & 0.000 & 0.239 & 0.000\\
 & Before Test 2 type A& 0.260 & 0.000 & 0.299 & 0.000 & 0.227 & 0.000 & 0.308 & 0.000 & 0.274 & 0.000\\
 & Before Test 2 type B & 0.251 & 0.000 & 0.307 & 0.000 & 0.226 & 0.000 & 0.325 & 0.000 & 0.272 & 0.000\\
 & Full Course type A& 0.240 & 0.000 & 0.275 & 0.000 & 0.255 & 0.000 & 0.265 & 0.000 & 0.255 & 0.000\\
 & Full Course type B & 0.213 & 0.000 & 0.291 & 0.000 & 0.203 & 0.000 & 0.303 & 0.000 & 0.211 & 0.000\\
 \hline
Java-2015 &  &  &  &  &  &  &  &  &  &  & \\
 & Before Test 1 type A& 0.230 & 0.001 & 0.274 & 0.000 & 0.253 & 0.000 & 0.242 & 0.001 & 0.231 & 0.001\\
 & Before Test 1 type B & 0.117 & 0.106 & 0.282 & 0.000 & 0.157 & 0.029 & 0.235 & 0.001 & 0.170 & 0.018\\
 & Before Test 2 type A& 0.227 & 0.001 & 0.261 & 0.000 & 0.281 & 0.000 & 0.210 & 0.003 & 0.174 & 0.015\\
 & Before Test 2 type B & 0.089 & 0.217 & 0.264 & 0.000 & 0.168 & 0.019 & 0.234 & 0.001 & 0.060 & 0.405\\
 & Full Course type A& 0.175 & 0.015 & 0.208 & 0.004 & 0.273 & 0.000 & 0.166 & 0.021 & 0.118 & 0.102\\
 & Full Course type B & 0.041 & 0.575 & 0.267 & 0.000 & 0.193 & 0.007 & 0.206 & 0.004 & -0.005 & 0.942\\
 \hline
Java-2016 &  &  &  &  &  &  &  &  &  &  & \\
 & Before Test 1 type A& 0.163 & 0.016 & 0.186 & 0.006 & 0.218 & 0.001 & 0.206 & 0.002 & 0.149 & 0.028\\
 & Before Test 1 type B & 0.009 & 0.899 & 0.210 & 0.002 & 0.067 & 0.328 & 0.207 & 0.002 & 0.020 & 0.764\\
 & Before Test 2 type A& 0.205 & 0.002 & 0.255 & 0.000 & 0.271 & 0.000 & 0.239 & 0.000 & 0.176 & 0.009\\
 & Before Test 2 type B & 0.058 & 0.392 & 0.261 & 0.000 & 0.144 & 0.034 & 0.246 & 0.000 & 0.040 & 0.560\\
 & Full Course type A& 0.184 & 0.006 & 0.244 & 0.000 & 0.253 & 0.000 & 0.239 & 0.000 & 0.176 & 0.009\\
 & Full Course type B & 0.064 & 0.349 & 0.237 & 0.000 & 0.178 & 0.008 & 0.235 & 0.000 & 0.042 & 0.536\\
 \hline
    \end{tabular}
    \caption{Spearman Correlation between Graph Attributes and Final Grade for different Graph Types}
    \label{tab:graph_corr}
\end{table*}

\subsection{RQ2. What features of students' study habits and social connections are most predictive of student performance?}

We used $\chi^2$ scores and random forests to identify the most informative features in each of the different timespans of these classes.  A set of features identified for DM-2013 is shown in Table \ref{tab:features_dm2013}. The $\chi^2$ feature ranking is based on testing for the statistical independence of the individual features, where features with a high degree of dependence are less likely to be selected. Random forest feature selection, by contrast generates trees using the Gini impurity of each feature to make selections and favors the features that provides the highest reduction for the tree on each step. Most of the features that were selected are common between both methods such as inconsistency, total time, and total actions. However, some features that were selected by the $\chi^2$ model such as betweenness centrality are not selected by the random forest model while others selected by the random forests such as degree are absent in $\chi^2$.  Further analysis however, showed that for most of the features that were selected by only one model, they were replaced by a different but highly correlated feature (0.8 or more).  We therefore focused solely on the $\chi^2$ features for the rest of our analysis.

For all of the classes we ranked the features and we were able to observe a sudden drop in the $\chi^2$.  We used these drops to select the cutoff point for each of the features in our predictive models.  Based upon those cutoff points for the  DM-2013 class, we decided to use 15 features for the models before tests 1 and 2, and 14 for the full-semester models, while for DM-2015, 15 features seemed more appropriate for all of the timeframes. For Java-2015 we kept 16 features for the before test 1 model, 14 features for the before test 2 model, and 17 features for the whole semester model. For Java-2016, we kept 13 features for the before test 1 model, 19 features for the before test 2 model, and 21 features for the whole semester model.

To answer our second research question, we then examined the top 15 features in detail to determine what elements were most informative across classes.  When we compare across the classes and timeframes we found that the most important features were:

\begin{itemize}
    \item The total time spent in both browser and study sessions
    \item The total number of actions performed in both browser and study sessions
    \item The number of study and browser sessions
    \item The number of homogeneous sessions among study and browser sessions
    \item The number of answers posted on Piazza.
\end{itemize}

These findings show that some study patterns, such as spending more time on the online class tools, performing more online actions, generating more sessions, focusing on one tool at a time per session, and answering more questions on Piazza are all distinguishing between high-performing and lower-performing students across these classes, even as some of the tools used varied across classes (e.g. DM classes use WebAssign for their assignments and Java classes use Github for their project submissions). The fact that most of these features were selected for both the browser and study sessions shows that these two types of sessions are informative and defining features based upon only one would not cover the variance of the data as much as they do when used in concert.  There are other features that were frequently among the top 15 of the different classes, while they might be missing in one or two timeframes, such as the Betweenness Centrality in the social graph which was the first chosen feature for most of the cases and Inconsistency of sessions. 

Prior studies of MOOCs have shown that while social graph features can be predictive of student performance, when compared to the other study habit features they cannot add much value to the predictive models \cite{gitinabard18, taylor14}. The fact that betweenness centrality in the social graph and the number of Piazza answers were almost always among the top predictive features in our models shows that in the blended courses being socially active does make a difference, separate from other study habits. While the social graph features are highly correlated, the selection of Betweenness Centrality shows that only actively participating in the forum, whether asking or answering questions, is not enough. Since these raw graphs are directed, a user with high betweenness centrality must have connections to and from many others, which means they have both received replies from, and replied to many people. The inconsistency measure represents the length of the gaps between the student's different sessions and how the total number of sessions they completed compares to the maximum count in the course. This feature measures how frequently or infrequently a student accessed the class material.  Thus a student with a high inconsistency score is one that typically goes offline from the course for a long period of time and has fewer total sessions than the others. We found that this measure had a weak negative correlation with student performance.

One key difference between the discussion forums in blended courses and MOOCs is the users' relative participation rate. Prior research on MOOCs has shown that the students' participation in the discussion forums is very low in the MOOCs. For example, in our prior work on MOOCs only 5\% of the students posted or replied on the class discussion forum, while for all the blended courses of this study, the participation rate was more than 60\%. While participation on Piazza was not mandatory for any of the classes, it was encouraged as the main source of advice. The low participation rate in MOOCs may explain why the students' social attributes were not as informative as their study habits, while in blended courses the social networks seem to be more helpful. One additional interesting point about the social network metrics is that both in-degree and out-degree were often not as important as the betweenness centrality score. As mentioned previously, the students with high betweenness centrality scores are the ones that have a privileged place within the network and link many of the users to each other. This shows that giving or receiving answers alone does not determine students' performance, while communicating both ways with more of the class members by having connections to a wider variety of people does.

\begin{table*}[]
    \centering
    \begin{tabular}{|lc|lc|lc|}
    \hline
    By Test 1 && By Test 2 && All&\\
    Feature & $\chi^2$&Feature & $\chi^2$&Feature & $\chi^2$\\
    \hline
    Betweenness & 1426.973 & Betweenness & 781.446 & Betweenness & 1831.403\\
Total Time Study & 256.560 & Total Time Study & 612.321 & Total Time Study & 741.372\\
Out Degree & 220.393 & Inconsistency Study & 372.337 & Inconsistency Study & 702.950\\
In Degree & 206.330 & Total Actions Browser & 346.036 & Out Degree & 491.802\\
Total Time Browser & 162.064 & Total Actions Study & 340.501 & In Degree & 489.007\\
Total Actions Study & 123.205 & Out Degree & 265.480 & Inconsistency Browser & 405.542\\
Piazza Answers Browser & 112.543 & Inconsistency Browser & 264.976 & Total Time Browser & 217.037\\
Total Actions Browser & 92.468 & Total Time Browser & 259.661 & NumSessions Browser & 177.335\\
Inconsistency Study & 78.795 & In Degree & 258.750 & Total Actions Browser & 173.722\\
Inconsistency Browser & 65.326 & Piazza Answers Browser & 112.543 & Homogeneous Browser & 164.077\\
NumSessions Browser & 35.667 & NumSessions Browser & 98.720 & Total Actions Study & 161.182\\
NumSessions Study & 31.152 & Homogeneous Browser & 87.455 & NumSessions Study & 153.776\\
Piazza Questions Browser & 30.508 & NumSessions Study & 79.349 & Homogeneous Study & 133.198\\
Homogeneous Browser & 30.375 & Homogeneous Study & 68.937 & Piazza Answers Browser & 112.543\\
Homogeneous Study & 27.201 & Piazza Questions Browser & 30.508 & Piazza Questions Browser & 30.508\\
Heterogeneous Browser & 7.323 & Heterogeneous Browser & 13.228 & Heterogeneous Study & 24.126\\
AvgActionsPerSession Study & 6.582 & Heterogeneous Study & 11.527 & AvgActionsPerSession Study & 20.880\\
Heterogeneous Study & 4.478 & AvgActionsPerSession Study & 7.862 & Heterogeneous Browser & 14.291\\
AvgActionsPerSession Browser & 3.840 & AvgDurationOfSession Study & 5.697 & AvgDurationOfSession Study & 12.508\\
\% Heterogeneous Browser & 3.117 & AvgActionsPerSession Browser & 3.353 & AvgActionsPerSession Browser & 8.667\\
Piazza Ratio Browser & 2.239 & AvgDurationOfSession Browser & 2.784 & AvgDurationOfSession Browser & 5.103\\
\% Heterogeneous Study & 2.171 & AvgTimeBwSessions Study & 2.153 & \% Heterogeneous Study & 3.015\\
AvgDurationOfSession Study & 1.637 & Piazza Ratio Browser & 1.878 & Piazza Ratio Browser & 2.361\\
Hub Score & 0.932 & AvgTimeBwSessions Browser & 1.220 & AvgTimeBwSessions Study & 2.215\\
Authority Score & 0.912 & \% Heterogeneous Browser & 1.084 & Hub Score & 1.455\\
AvgTimeBwSessions Study & 0.784 & \% Heterogeneous Study & 1.083 & Authority Score & 1.445\\
AvgTimeBwSessions Browser & 0.488 & Authority Score & 0.999 & AvgTimeBwSessions Browser & 0.979\\
\% Homogeneous Study & 0.101 & Hub Score & 0.562 & \% Heterogeneous Browser & 0.402\\
\% Homogeneous Browser & 0.091 & \% Homogeneous Browser & 0.196 & \% Homogeneous Browser & 0.291\\
AvgDurationOfSession Browser & 0.001 & \% Homogeneous Study & 0.166 & \% Homogeneous Study & 0.077\\
\hline
    \end{tabular}
    \caption{Feature Rankings  DM-2013}
    \label{tab:features_dm2013}
\end{table*}

\subsection{RQ3. How early can we predict students' performance in these classes using the data from the same class?}

The results for each of the different algorithms on different classes are shown in Table \ref{tab:res_same_class}. As the table shows, in most cases using the data from the whole semester allowed us to identify the distinction students with an F1 score of 70\% or more. We believe that the reason why whole semester data outperforms the smaller timeframes across the semester is that the additional information helps to account for variance among the student groups.  While the students' behaviors may not be similar in specific timeframes of the semester, the overall habits of the distinction group seem to be similar across classes. If we truncate our data to an earlier timeframe such as before test 1 or before test 2, the performance of the models is reduces, but still, most of them are able to achieve an F1 score of about 60\%.

\begin{table}[h]
    \centering
    \begin{tabular}{|lccc|}
    \hline
   Algorithm & Before Test 1 & Before Test 2 & Full Data\\
   \hline
 DM-2013 &  &  & \\
SVM & 0.540 & 0.568 & 0.619\\
Random Forest & 0.527 & 0.559 & 0.674\\
Logistic Regression &\textbf{ 0.638 }& \textbf{0.615} & \textbf{0.703}\\
\hline
DM-2015 &  &  & \\
SVM & \textbf{0.729} & \textbf{0.744} & \textbf{0.744}\\
Random Forest & 0.661 & 0.730 & 0.705\\
Logistic Regression & 0.639 & 0.680 & \textbf{0.745}\\
\hline
Java-2015 &  &  & \\
SVM & 0.483 & 0.541 & 0.646\\
Random Forest & \textbf{0.582} & 0.443 & \textbf{0.706}\\
Logistic Regression & \textbf{0.583} & \textbf{0.559} & 0.573\\
\hline
Java-2016 &  &  & \\
SVM & 0.775 & 0.779 & 0.801\\
Random Forest & 0.740 & \textbf{0.792} & \textbf{0.831}\\
Logistic Regression & \textbf{0.785} & 0.752 & 0.785\\
\hline
    \end{tabular}
    \caption{F1 Scores for Same Class Predictions}
    \label{tab:res_same_class}
\end{table}

For our third research question, we wanted to find out how early we can cut the data and make performance predictions. Our findings show that when using the early stage data, as early as the first or second midterm of the classes, while not as accurate as the whole semester data, we are still able to predict the students' performance with reasonable accuracy in most of the classes. Thus, our next step would be to analyze whether we can train our models on a prior offering of these courses and then test them on an early stage of the class. 

\subsection{RQ4. Will predictive models generated from one offering of a course transfer to another offering of the same course?}

In order to be able to use the models for making predictions, we need them trained before we have final class grades. Thus, it makes the models more useful if we are able to train them on one offering of a class and apply them to another class which is still ongoing. In our present work in order to make cross predictions across offerings, we selected the best performing algorithm for each timespan of each course. As a result, for DM-2013 for example, we trained a logistic regression using 15 features for before test 1 and before test 2 timespans and 14 features for the whole semester. 

Prior research has shown that while using a whole semester of data from the earlier offering might provide more information for the training of a model, using the data from the same timespan of the prior offering may produce a more generalizable model \cite{gitinabard18}. Thus, when making predictions for each timespan we used two different models. The first was trained on the same timeslice of the prior offering while the second model is trained on the whole semester data of the earlier offering.  

The cross offering results are shown in Table \ref{tab:res_cross_offering}. As these results show, using the whole semester data of the first offering provides us with a better predictive model on early stages of the second class. To answer our fourth research question, our findings indicate that despite the differences in the top features across offerings, these models are able to predict the distinction group in the second course offering with an F1 score of 60\% or more before the first exam in the class.
\begin{table}[h]
    \centering
    \begin{tabular}{|lcc|}
    \hline
   Data Trained On & Before Test 1 & Before Test 2\\
   & Second Offering & Second Offering\\
   \hline
 DM-2013 Before Test 1 & 0.629 & \\
 DM-2013 Before Test 2 &  & 0.636\\
 DM-2013 All Data & 0.636 & 0.652\\
Java 15 Before Test 1 & 0.548 & \\
Java 15 Before Test 2 &  & 0.717\\
Java 15 All Data & 0.672 & 0.746\\
   \hline
    \end{tabular}
    \caption{F1 Scores for Cross Offering Predictions}
    \label{tab:res_cross_offering}
\end{table}

\subsection{RQ5. Will prediction models generated on one course transfer to another?}
Finally, in order to use the models for early prediction they must be trained before the class is over. Using models from other courses can be beneficial in cases where a course is being offered for the first time, thus we need to study whether the predictive models trained on one class can be used to make predictions on a later offering of a different class. In this case, as in the cross offering predictions, we use both data from the same timespan and data from the whole semester of the earlier course to make predictions on early stages of the latter. Here we tried the model trained on  DM-2013 on Java-2015 and Java-2016. While DM-2015 and Java-2015 were offered in the same semester, we also tested the Java-2015 model on the performance in DM-2015 so that our findings are not only based on models trained on DM classes and we can also see how well the models trained on a Java class can perform on a DM offering.

Our findings are shown in Table \ref{tab:res_cross_course}. As these results indicate, as with the cross offering models, using the data from the whole semester of the earlier class seems to produce a better-trained model. To answer our fifth research question, these models were able to make predictions on the later offering of another course with an F1 score of 60\% or more, even when using only the data collected before the first test.
\begin{table*}[]
    \centering
    \begin{tabular}{|lcccccc|}
    \hline
Data Trained on & DM-2015  & Java-2015  & Java-2016  & DM-2015  & Java-2015  & Java-2016 \\
 & Before Test 1 & Before Test 1 & Before Test 1 & Before Test 2 & Before Test 2 & Before Test 2\\
 \hline
 DM-2013 Before Test 1 &  & 0.628 & 0.655 &  &  & \\
 DM-2013 Before Test 2 &  &  &  &  & 0.543 & 0.610\\
 DM-2013 All Data &  & 0.632 & 0.655 &  & 0.571 & 0.664\\
\hline
Java-2015 Before Test 1 & 0.543 &  &  &  &  & \\
Java-2015 Before Test 2 &  &  &  & 0.628 &  & \\
Java-2015 All Data & 0.709 &  &  & 0.698 &  & \\
   \hline
    \end{tabular}
    \caption{F1 Scores for Cross Course Predictions}
    \label{tab:res_cross_course}
\end{table*}

\subsection{RQ6. How will these models perform in  identifying at-risk students?}

While the number of at-risk students in our classes were low, we decided to train and test our prediction models for identifying at-risk students as well. The classes included in this study are C-wall classes, meaning the students need a C or better grade to proceed to the further courses in Computer Science curriculum. Thus, we defined all the students gaining a lower than C grade as at-risk students. There were between 20 and 35 at-risk students in these classes and to prevent our models from fitting to the majority class, we sampled the not-at-risk students in our data to twice the number of at-risk students.

The results of at-risk prediction across the classes are shown in Table \ref{tab:at-risk}. As these results show, despite the fewer samples remaining in the at-risk prediction, the models are still performing well and sometimes better than the distinction prediction.

\begin{table}[h]
    \centering
    \begin{adjustbox}{width=0.48\textwidth}
    \begin{tabular}{|lcccc|}
    \hline
Trained on & Test Data & DM 2015  & Java 2015 & Java 2016\\
\hline
DM 2013 & Before test 1 & 0.77 & 0.74 & 0.79\\
Java 2015 & Before test 1 & 0.75 & - & 0.81\\
DM 2013 & Before test 2 & 0.78 & 0.78 & 0.78\\
Java 2015 & Before test 2 & 0.64 & - & 0.84\\
\hline
\end{tabular}
\end{adjustbox}
    \caption{F1 Scores for Cross Class At-risk Classification}
    \label{tab:at-risk}
    
\end{table}
\section{Conclusions}
In this study, we first analyzed two different methods for generating social graphs from the students' interactions on classroom discussion forums. We found that, as in MOOCs, when the users are able to see all of the replies on a thread without clicking on them, assuming that the students read all the replies before posting a new one gives us a more informative social network. We also defined several features based on students' study behavior and social interactions and used those features to make predictions of the students' class performance. Our findings showed that in contrast to prior studies of MOOCs, the students' social metrics seem to be as or more helpful for predicting student performance than the students' study behaviors. We also found that when using a suitable prediction algorithm, the defined features can predict the distinction group students with an F1 score of about 60\% even before completing the first test. Our results also show that these predictive models were able to generalize from earlier offerings of one course to later offerings of the same course or even to other courses. Our results also showed that these models are able to be trained to predict at-risk students in these classes, even considering the low number of students in this group. These generalized models show that, as we hypothesized, most of the behaviors and social interaction metrics of higher performing students are similar across different classes. We also examined these metrics specifically for the students who were enrolled in both DM-2015 and Java-2015 (126 students) and found that there was a positive but weak correlation (0.3 - 0.4) between most of the metrics of these students in different classes.  The fact that correlation is positive shows that the metrics are mostly consistent for the same student in different classes.  We attribute the weakness of the correlations to structural differences between the courses and effect of other unshared students who do affect the metrics.   Moreover, the importance of some of the graph features shows that, consistent with social learning theory and the theory of Communities of Practice, being involved in the discussions is important to the students' performance. While prior research does not imply and we did not assume that more connections are necessarily better, the selection of graph metrics among the most predictive features suggests that at least in our cases, the students who were more engaged in the discussions performed better in the class. These cross class models can help the instructors make predictions on their students' performance as early as the first test and intervene by making more support or resources available for the lower performing students. The results of this study can also be used to help in identifying harmful study patterns such as inconsistency. 


One of the limitations of this work was that we did not have access to the WebAssign data from the DM-2015 class. While our 2013 data suggests that the majority of student actions were WebAssign transactions, the only features reflecting these actions that were considered to be important were the session features. This lack of data seems to have a relatively small effect on the generality of the model across classes since this data source is missing for all the students and thus everyone's sessions will be shorter.

In future, we plan to expand our analysis by using data from more offerings of these courses. We also plan to study the post and reply content as well as commit messages generated by the students and define some text-based features based on those to make our models more accurate. The students' teamwork behavior and their participation in the course projects can also provide us with more insight on their work and study habits which we can use to identify students in need of support and provide tailored guidance. We are in the process of planning course interventions but we have not carried any out at this time. We plan to use these predictive models to identify, and intervene with at-risk students early in the semester and to provide them with more resources such as group study and peer tutoring.

\bibliographystyle{abbrv}
\bibliography{references}

\end{document}